\documentclass{desyproc}

\begin{document}
\title{The Higgs Physics Programme \\
 at the International Linear Collider}

\author{{\slshape Felix Sefkow$^1$}\\[1ex]
$^1$DESY, Notketra{\ss}e 85, 22607 Hamburg, Germany }

\contribID{xy}

\confID{8648}  
\desyproc{DESY-PROC-2014-04}
\acronym{PANIC14} 
\doi  

\maketitle

\begin{abstract}
The talk summarises the case for Higgs physics in $e^+e^-$ collisions 
and explains how Higgs parameters can be extracted in a model-independent way 
at the International Linear Collider (ILC). 
The expected precision will be discussed in the context of projections for the
experiments at the Large Hadron Collider (LHC). 
\end{abstract}

\section{Introduction}

The discovery of a Higgs boson, honoured with the 2013 Nobel prize in physics, marks a turning point in particle physics, 
as the last missing building block of the Standard Model falls into place and opens the door to completely new studies of a 
particle unlike every other discovered before. 
Like in many earlier instances in the history of particle physics, it did not come as a surprise, but was anticipated and sought for. The Higgs mass had been predicted with increasing precision from the analysis of electro-weak quantum corrections, in which measurements at the previous generation of  $e^+e^-$ colliders played a prominent r\^ole. 

Today, Higgs physics has been identified as one of the prime "drivers" of the field, as a compelling line of research with great promise, where surprises may be expected. 
The main question is to fully establish the profile of the Higgs particle, measure its quantum numbers and, above all, its precisely predicted couplings to almost all other fundmental particles, and to find out whether it fulfils its r\^ole in the Standard Model, or whether it holds the key to new physics beyond. 

The accuracy, which is required in order to detect possible mechanisms behind electroweak symmetry breaking through deviations of the Higgs couplings from their pure Standard Model values, has been quantitatively investigated in the framework of the Snowmass study 2013~\cite{snowmasshiggs}. 
Popular models like two-Higgs doublet or composite Higgs schemes, which predict new particles at the TeV scale, and which are still compatible with recent limits from direct searches at the LHC, typically lead to such deviations in the per-cent or sub-percent range. 
This sets the scale of the future experimental challenges and demonstrates the discovery potential of precision measurements in the Higgs sector. 

\subsection*{The ILC and its detectors}
 
 The ILC has been proposed as the next big high energy accelerator project. 
 It is designed to have centre-of-mass energies ranging from 250 to 500~GeV and is upgradeable to reach 1~TeV. 
 The delivered luminosity increases with energy and amounts to typically 100 -- 300~fb$^{-1}/$y, 
 with beam polarisations of up to 80\% and 30\% for electrons and positrons, respectively. 
 The superconducting technology is mature, as is demonstrated by the on-going construction of the European XFEL at DESY, which uses a very similar design at industrial scales. 
 A technical design report (TDR)~\cite{tdr} for the ILC has been completed in 2012, a proposed site has been selected in the Kitakami mountains in the North of Japan, and the project is currently being discussed at ministerial levels. 

Two detector concepts have been proposed~\cite{dbd} for the ILC, which have been optimised for precision, as radiation hardness and rate capability requirements are very relaxed with respect to those at the LHC.  
The detectors feature highly granular and compact calorimeters for particle flow reconstruction, ultra-thin and precise trackers, and vertex detectors capable of identifying not only beauty but also charm quarks. 
Detailed designs have been implemented in the simulations to evaluate the physics potential under realistic conditions, including beam-induced backgrounds.

\section{Measurements of Higgs couplings}

It is instructive to recall the necessary ingredients to a measurement of a coupling strength. 
The number of particles $N$ observed in a given final state $f$, normalised to integrated Luminosity $L$, is given by the product of cross-section $\sigma$ and branching fraction $\cal B$, which is the ratio of partial width $\Gamma_f$ to total width 
$\Gamma_T$.
The couplings to the initial and final state, $g_i$ and $g_f$, enter via the production cross section and the partial width, such that one has
\begin{equation}
N / L = \sigma \cdot {\cal B} = \sigma \cdot \Gamma_f / \Gamma_T \sim g_i^2 \cdot g_f^2\, / \,  \Gamma_T \, .
\end{equation}
In order to extract $g_f$, one needs a measurement of the inclusive cross section -- to obtain $g_i$ -- and the total width. 
In the Z line shape analysis at LEP, the width of the Z resonance was directly observable, and the cross section 
in the $e^+e^-$ final (and initial) state provided a normalisation of the couplings of the Z to fermions. 
The width of the Higgs particle, however, is expected to be about 4~MeV in the Standard Model and too narrow to be resolved experimentally, so it has to be extracted from the branching ratio of a channel, for which the coupling is already known, e.g.\ from a production measurement, 
\begin{equation}\label{Eq:GammaT}
\Gamma_T = {\cal B} / \Gamma_f \sim {\cal B} / g^2_f
\end{equation}
At the LHC the total cross section and total width are poorly constraint, and in general the Standard Model values are assumed. 
At the ILC, however, one can make use of the unique features of an $e^+e^-$ collider to obtain a self-contained set of observables.

\subsection{Higgs production at the ILC}
\begin{figure}[htb]
\begin{center}
\includegraphics[width=0.25\textwidth]{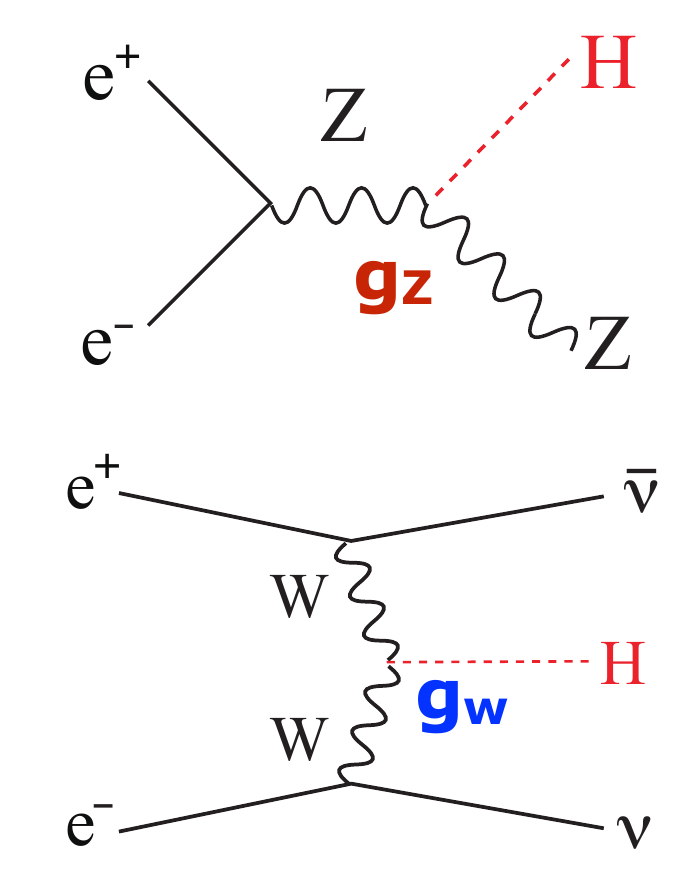}
\hspace{1cm}
\includegraphics[width=0.35\textwidth]{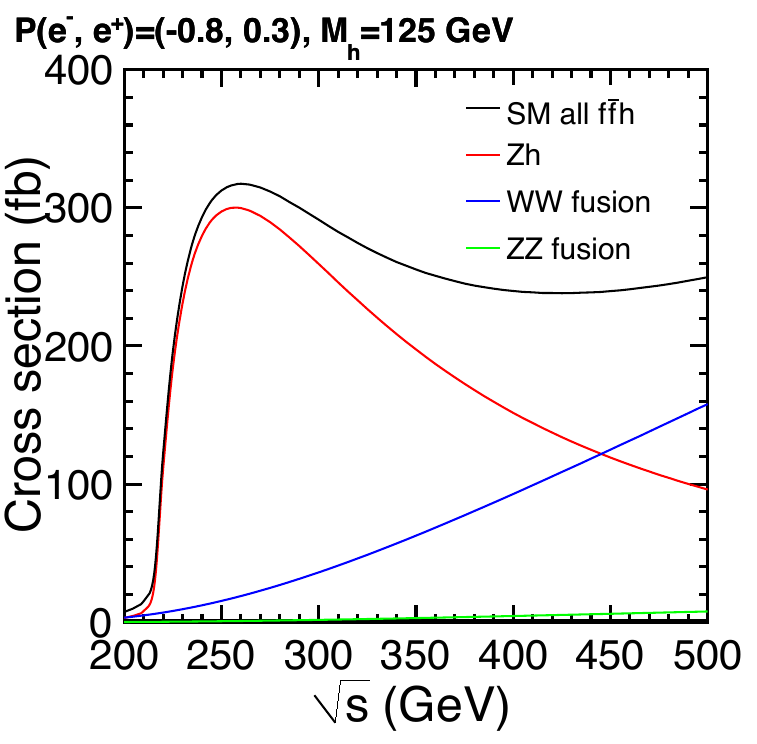}
\end{center}
\caption{Higgs production diagrams and cross section vs.\ centre-of-mass energy.}\label{Fig:Hprod}
\end{figure}

The dominant Higgs production processes at the ILC are Higgs strahlung and W fusion.~\cite{ilctdrphys}.
Figure~\ref{Fig:Hprod} shows the diagrams and the dependence of the cross-section on the centre-of-mass energy.
Higgs strahlung as an $s$ channel process dominates at threshold, whilst the cross section of the $t$ channel process W fusion increases logarithmically with energy and takes over at about 450~GeV. 
Here, one has made use of the beam polarisation to enhance the cross section. 
Now, since at an $e^+e^-$ machine one can control the energy of the incoming fermions, one can select the dominant process by tuning the beam energy. 

\begin{figure}[htb]
\begin{center}
\includegraphics[width=0.35\textwidth]{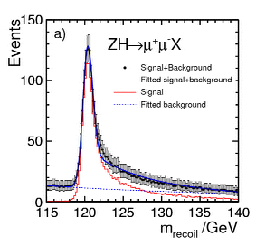}
\hspace{1cm}
\includegraphics[width=0.45\textwidth]{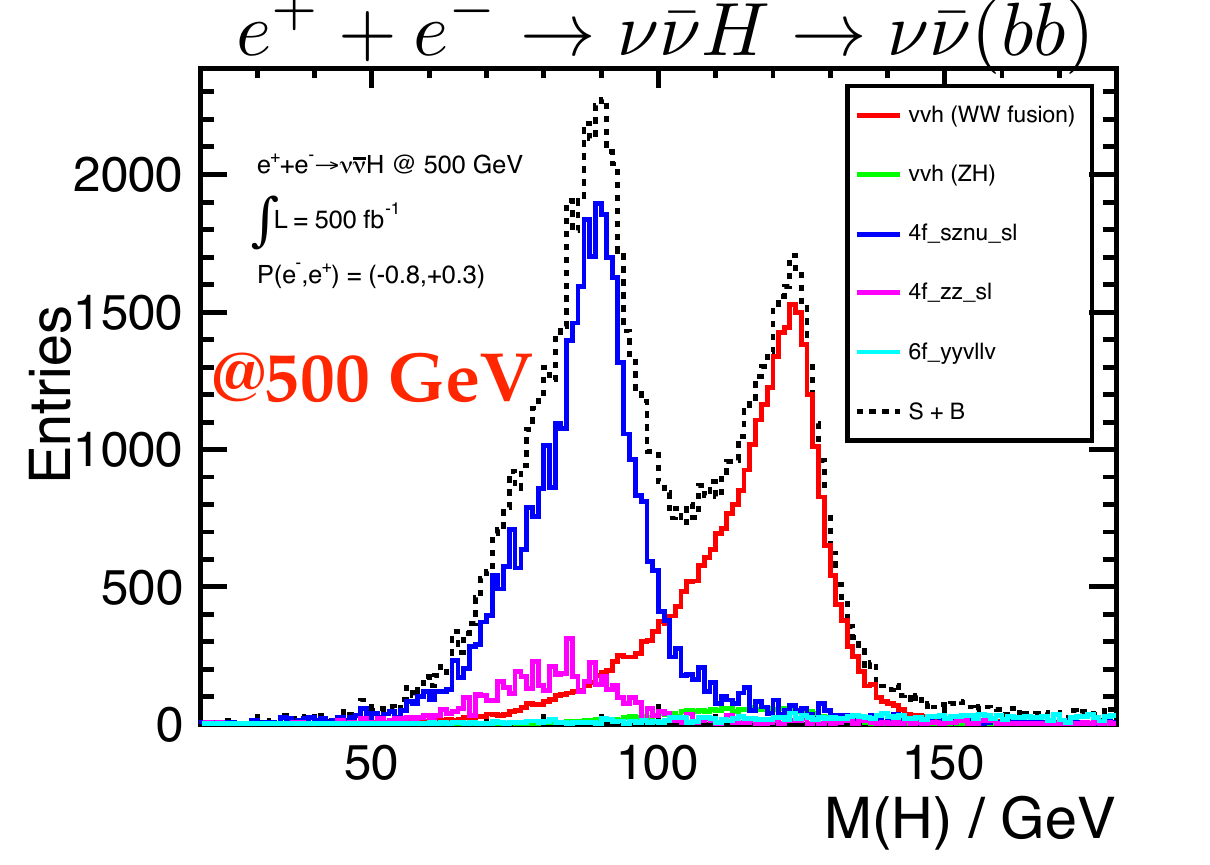}
\end{center}
\caption{Higgs signal in the recoil mass spectrum (ZH production), and in the $b\bar{b}$ di-jet mass (W fusion).}\label{Fig:Hrecoil}
\end{figure}

Another consequence of the well-defined initial state is the possibility to apply kinematic constraints. 
In ZH events, a Higgs signal can be observed in the spectrum of recoil masses against the Z decay products,
\begin{equation*}
M_{\mbox{\small recoil}}^2 = E^2 - p^2 \: \mbox{with} \:  E = \sqrt{s} - E_Z \;\mbox{and} \; p=p_Z
\end{equation*}
 This works best for Z decays into muon pairs, as shown in Figure~\ref{Fig:Hrecoil}, but also well for the electron channel, whilst for hadronic Z decays it is more difficult. 
Here, no requirements whatsoever on the Higgs final state have been made, it can even be invisible, and thus the measurement is fully inclusive.
It provides an absolute normalisation for all branching ratios into specific final states and a model-independent extraction of 
the absolute value of $g_Z$, the Higgs Z coupling, which is {\em the} central measurement of the Higgs coupling analyses. 

\subsection{The Higgs total width}

The Higgs mass of 125~GeV is almost ideally suited for the study of a large number of decay modes with not too small branching ratios. 
However, the fraction of decays into Z pairs is only a few per-cent and the statistics for specific Z channels very small.
An extraction of the total width, using Eq.~\ref{Eq:GammaT} with $g_Z$ and ${\cal B}({\rm H}\rightarrow {\rm ZZ}^*)$ is in principle possible, but would suffer from large uncertainties of $\sim 20\%$.

It is more advantageous to use the W fusion cross section and the branching ratio ${\cal B}({\rm H}\rightarrow {\rm WW}^*)$.
Since in W fusion the Higgs is accompanied by two neutrinos, the recoil method cannot be applied for a decay-mode independent measurement, but a specific Higgs channel must be used. 
Both the $b\bar{b}$ and the $WW^*$ channel are suited~\cite{duerig}; the $b\bar{b}$ signal is shown in Figure~\ref{Fig:Hrecoil}.
Since these decay modes are also measured in HZ production, the ratio $g_W/g_Z$ and thus $g_W$ can be extracted and 
$\Gamma_T$ from Eq.~\ref{Eq:GammaT}. 
Now one has all ingredients to convert also the other branching ratio measurements into absolute couplings, 

\subsection{Higgs couplings to fermions and the self-coupling}

Thanks to the relatively benign beam conditions at the ILC vertex detector systems can be realised which can not only identify 
$b$ flavoured hadron decays on the basis of the finite decay length, but can also tag charmed hadrons and disentangle prompt open charm from tertiary vertices, which originate from $b\rightarrow c$ decays. 
Particularly well suited are ZH events with Z decaying into neutrinos, such that the final state consists of the two jets from the Higgs only, giving a signal in the diet invariant mass. 
A multivariate analysis of the vertex topologies then yields a simultaneous measurement of 
${\cal B}({\rm H}\rightarrow b\bar{b})$, ${\cal B}({\rm H}\rightarrow c\bar{c})$ and ${\cal B}({\rm H}\rightarrow gg)$, and thus 
$g_b$, $g_c$ and a model-dependent value for $g_t$, like the $\gamma\gamma$ mode.

The measurement of the coupling to the second quark generation is unique for testing the mass dependence of the Higgs coupling in the quark sector, since couplings to $u$, $d$ and $s$ quarks are unobservable. 
In the lepton sector, $g_{\tau}$ can be measured well, but in the H$\rightarrow\mu\mu$ channel only very few events can be observed and only at the highest energies attainable at the ILC, where luminosity and cross section are maximal. 

The direct observation of the top Higgs Yukawa coupling is made though a production cross section measurement for the 
$t\bar{t}$H channel, where, e.g., a Higgs is radiated from one of the two quarks in a $t\bar{t}$ pair.  
This involves the analysis of complex 8 or 10 fermion final states, where eben after using flavour tags and di-jet masses, the signal basically consists of an excess over expectation without $t\bar{t}$H coupling. 
This is a particularly good example for cases where a large gain in precision can be obtained from a combined evaluation of ILC and LHC data, see below. 

Finally, a measurement of the Higgs self-coupling would represent the last cornerstone in establishing the Higgs profile and demonstrating that it has the properties required for electro-weak symmetry breaking. 
The strength $g_{HHH}$ can be measured at the ILC, albeit with only moderate precision. 
This is due to the fact that ZHH events are not only produced with diagrams involving triple-Higgs coupling, but also through processes like double Higgs strahlung, which constitute an irreducible background. 
The situation is more favourable in the case of W fusion leading to $\nu\bar{\nu}$HH events, therefore the best precision is obtained at highest energies, where the dilution is less and luminosity and cross section are largest. 

\section{Global fits and achievable precision: Summary}

\begin{figure}[htb]
\begin{center}
\includegraphics[width=0.40\textwidth]{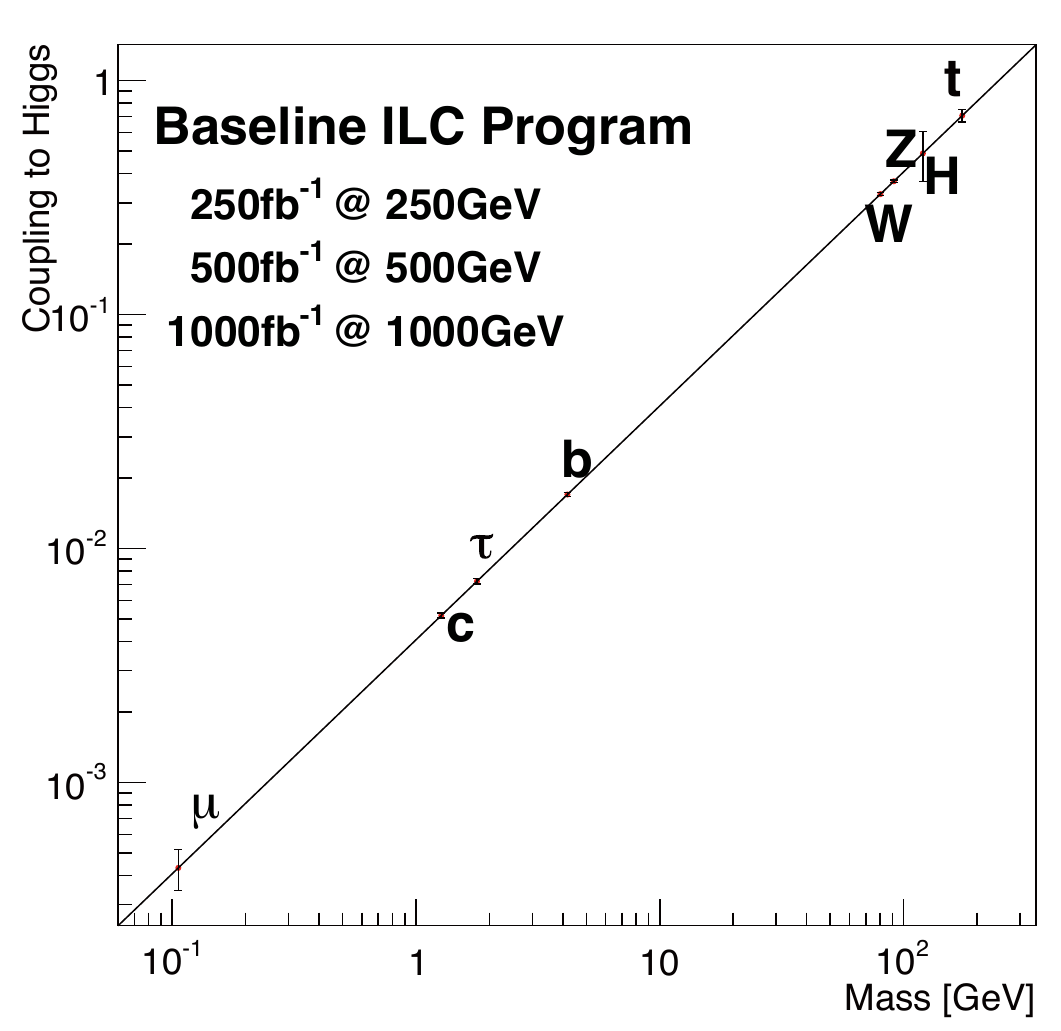}
\hspace{1cm}
\includegraphics[width=0.50\textwidth]{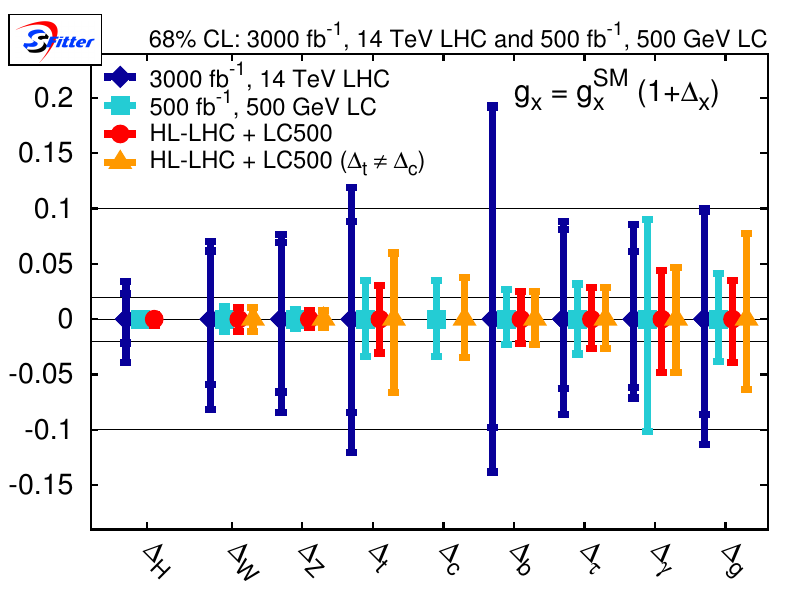}
\end{center}
\caption{Higgs coupling strengts, mesurerd at the ILC, as a function of mass: relative precision for expected ILC and LHC results, including the luminosity upgrade, and combination of data. }\label{Fig:Hprecision}
\end{figure}

In a staged running scenario, each centre-of-mass energy, 250, 500 and 1000~GeV, provides an independent set of measurements.
Altogether, 33 measurements of $\sigma\cdot {\cal B}$ values are made and injected into a global fit with 10 free parameters  -- the couplings to W, Z and $t$, $b$, $c$, $\tau$, $\mu$ fermions, indirect to $gg$, $\gamma\gamma$ pairs, and the total width 
$\Gamma_T$. 
The result is shown in Figure~\ref{Fig:Hprecision}. 

The precision has been compared to that expected for the LHC~\cite{peskin} and its high-luminosity upgrade~\cite{zerwas}.
In these studies consistent assumptions and constraints have been used for both colliders' data sets, which is important for a fair comparison. 
As the Figure~\ref{Fig:Hprecision} shows, with linear collider results the per-cent and sub-per-cent level precision can be reached, which is required to detect deviations from the Standard Model in the magnitude expected in theories for mechanisms behind electro-weak symmetry breaking. 

\section{Bibliography}
 

\begin{footnotesize}



%

\end{footnotesize}


\end{document}